# Angularly quantized spin rotations in hexagonal LuMnO$_3$

Seung Kim[1], Jiyeon Nam[1],
Xianghan Xu[2], Sang-Wook Cheong[2], and In-Sang Yang[1,*]

[1]*Department of Physics, Ewha Womans University, Seoul Korea*
[2]*Rutgers Center for Emergent Materials and Department of Physics and Astronomy, Rutgers University, Piscataway, New Jersey, USA*

**ABSTRACT**

Optical control of the spin degree of freedom is often desired in application of the spin technology. Here we report spin-rotational excitations observed through inelastic light scattering of the hexagonal LuMnO$_3$ in the antiferromagnetically (AFM) ordered state. We propose a model based on the spin-spin interaction Hamiltonian associated with the spin rotation of the Mn ions, and find that the spin rotations are angularly quantized by 60, 120, and 180 degrees. Angular quantization is considered to be a consequence of the symmetry of the triangular lattice of the Mn-ion plane in the hexagonal LuMnO$_3$. These angularly-quantized spin excitations may be pictured as isolated flat bubbles in the sea of the ground state, which may lead to high-density information storage if applied to spin devices. Optically pumped and detected spin-excitation bubbles would bring about the advanced technology of optical control of the spin degree of freedom in multiferroic materials.

\* Corresponding author, e-mail: yang@ewha.ac.kr

INTRODUCTION

Application of spin control is looking forward to expanding its horizon to unprecedented level of data processing and storage capability [1, 2]. Understanding the spin-spin interaction in magnetic materials are important and required in order to control the spin degree of freedom in the materials. However, most investigations trying to discover the origin of magnetic properties have been encountered with mixed information of unknown origin, which make it hard to analyze the genuine response from the magnetic system under investigation.

Despite the aforementioned difficulty, magnetic technologies have been actively explored in various fields to overcome the practical issues with the $R$MnO$_3$ ($R$ = rare earths, Y, and Sc) which are well known materials possessing ferroelectric and antiferromagnetic (AFM) transitions simultaneously. This binary character has been investigated by several optical and magnetic techniques as well as theoretical studies because of its immense potential for application [3-7]. Most $R$MnO$_3$ materials exhibit a new possibility of tuning the response of the electric parameters to a magnetic field and vice versa, due to strong coupling of electric and magnetic order parameters [8-13].

The crystal structure of hexagonal $R$MnO$_3$ in P6$_3$cm space group leads to split the 3$d$ energy levels of Mn$^{3+}$-ions, due to change in the crystal field symmetry below the Curie-Wiess temperature T$_C$ (> 700 K) [14-17]. A Mn$^{3+}$-ion has four electrons in 3$d$ orbital (total S = 2). When a proper energy is supplied to the system, one of the electrons in $d_{x^2-y^2}/d_{xy}$ levels can be excited to the $d_{3z^2-r^2}$ level, called Mn $d$-$d$ transition [11, 16-18].

The Mn$^{3+}$-ions are placed at x = $\frac{1}{3}$ position of the triangular lattice in a $xy$ plane in paramagnetic phase. However, below Néel temperature $T_N$ (< 100 K), they move out of the ideal site due to Mn trimerization [15, 19-21], giving rise to two different spin-spin interaction integrals, intra-trimer interaction $J_1$ and inter-trimer interaction $J_2$.

Raman spectroscopy is based on the inelastic scattering related to electronic transitions in a material. Raman selection rule can help establish the excitations of the magnetic origin with high resolution [22, 23]. Previous Raman studies show that the spin excitations strongly correlated with a particular electronic transition are observed in hexagonal $R$MnO$_3$ below the $T_N$ [24-28]. Spin excitations of relatively high energy (~0.1 eV) are optically excited in the hexagonal $R$MnO$_3$ system through the resonance with the Mn $d$-$d$ transition [24, 28-30]. These are not phonons at the zone boundaries that could fold in due to structural or magnetic ordering.

LuMnO$_3$ is of the popular hexagonal manganite family crystalized in the P6$_3$cm group below $T_N$ [6, 14, 20, 31]. Unlike other rare-earth hexagonal $R$MnO$_3$, only Mn$^{3+}$-ion has spins in hexagonal LuMnO$_3$ [10, 32], and the spin excitations in the Mn-ions can be well separated without screening or overlapped effects by other than Mn-ion spins. Diverse articles have been reported regarding this material for the latest several years, focusing on the ensemble of the lattice constant, crystal structure, spin structure, phonon, magnon and spin exchange integrals [9, 12, 13, 15].

In this study, we present a microscopic explanation for the origin of spin excitations based on a simple spin-spin interaction Hamiltonian in hexagonal LuMnO$_3$ system. We propose a model associated with the spin rotation of the Mn ions in the symmetry of the triangular lattice with the AFM ordering. Our model can explain the spin excitation peaks observed in LuMnO$_3$ within the framework of the spin-spin interaction in the Mn-trimer network below $T_N$.

**RESULTS**

Figure 1 shows temperature-dependent Raman spectra of the hexagonal LuMnO$_3$ single crystal in cross polarization scattering geometry from 120 to 1050 cm$^{-1}$ range as a function of temperature. All the spectra were normalized by the intensity of the A$_1$ phonon at ~680 cm$^{-1}$. The A$_1$ phonon should be forbidden in the cross polarization, except in the resonance condition. Other A$_1$ (~266, 475 cm$^{-1}$), E$_1$ (~640 cm$^{-1}$), and E$_2$ (~318, 350, 461 cm$^{-1}$) phonon modes observed in the spectra are consistent with previous results for LuMnO$_3$ [12]. Besides these phonons, several broad peaks (~197, 580, 805 cm$^{-1}$) are prominent at low temperatures. These broad peaks disappear above a critical temperature ~80 K. (See the Supplementary Information.) As in various hexagonal $R$MnO$_3$, AFM spin ordering appears under ~100 K, and the T$_N$ values for LuMnO$_3$ suggested by previous papers are consistent with the critical temperature ~80 K [24, 33, 34]. This strongly suggests that these broad peaks are in harmony with the AFM spin ordering in LuMnO$_3$. In the sense that the peaks in the Raman scattering spectra measure the energy difference between the ground state and the excited state, it is reasonable to assume that the broad peaks represent the energy difference between the ground state and several excited states of the AFM spin ordering. The peak at 805 cm$^{-1}$ may be considered as the spin-flip excitation energy for the Mn$^{3+}$ ions in one triangle of the lattice (trimer) in LuMnO$_3$ just as the 760 cm$^{-1}$ peak in HoMnO$_3$ is claimed to be due to the spin-flip excitation [28]. What about the origin of the other broad peaks at lower wavenumbers? In order to answer the question, we first need to look for the physical parameters from the relationship between the measured Raman peak at 805 cm$^{-1}$ and the Hamiltonian for the spin-flip excitation of the Mn$^{3+}$ ions in the trimer.

The spin excitation peaks in hexagonal $R$MnO$_3$ have been observed by red lasers only [24, 29, 30]. The optical conductivity measurements on hexagonal $R$MnO$_3$ provide convincing evidence that the Mn *d-d* transition take place 1.5 ~ 1.8 eV at room temperature [11, 16-18]. In LuMnO$_3$, Mn *d-d* transition would occur around 1.615 eV at 300 K, and the Mn *d-d* transition peak blue shifts about 0.15 eV at 10 K [16]. Red excitation lasers of 620 ~ 700 nm wavelength supply the energies corresponding to 1.77 ~ 2 eV, close to the energies of Mn *d-d* transition of LuMnO$_3$ at 10 K. Many Raman scattering studies on hexagonal $R$MnO$_3$ system support our interpretation of resonance with Mn *d-d* transition well [24, 29, 30]. The resonance Raman scattering is specific only to the Mn *d-d* transition, so that the spin excitations observed by the red lasers are independent of the disturbance from the excitations of the $R$ ions. Therefore, the resonance Raman scattering would open a new opportunity to study the magnetic properties associated with the Mn ions selectively in hexagonal $R$MnO$_3$ system.

The resonance effect with the Mn *d-d* transition further supports that these broad peaks observed in LuMnO$_3$ below T$_N$ may be from the excitation of the Mn-ions. But are they due to the excitation in the spin structure of Mn-ions? Let us consider possible excited Mn$^{3+}$-ion spin configurations in triangular lattice within the AFM ordering symmetry. Our aim is to calculate the excitation energy ΔE, and match with the energies of the broad peaks observed in the Raman spectra.

Mn *d-d* transition would induce a transient excited state of different spin symmetry. The new transient spin-ordered state should be consistent with the AFM structure of the P6$_3$*cm* space group. Spin structures in this manuscript are described by unidimensional Γ representations as in diverse literatures [19, 21, 31]. Among the representations, Γ$_4$ is the most plausible candidate for the ground state of LuMnO$_3$ as displayed in Fig. 2a [20, 34, 35]. The Mn-ion spins are arranged on a trimer with counterclockwise configuration in the $z = c/2$ plane (gray solid spheres), while those are arranged with clockwise configuration in the $z = 0$ plane (black solid spheres). Layered MnO$_5$ bipyramids are separated far enough to ignore the interaction of the Mn-ions between the planes, so the main concern would be interactions in one *xy* plane [32, 35]. In this manuscript, we will assume that three Mn ions of a trimer in the $z = 0$ plane only are excited by the resonant light.

Raman selection rule is resulted from the conservation of total angular momentum, $\vec{J} = \vec{L} + \vec{S}$. Here, $\vec{J}$ is total angular momentum, $\vec{L}$ is orbital angular momentum, and $\vec{S}$ is spin angular momentum. Raman scattering has a two-photon process which should satisfy $\Delta J = 0$ or $\pm 2$. As was explained above, resonance Raman scattering in hexagonal *R*MnO$_3$ is linked with Mn *d-d* transition, so $\Delta L$ would remain unchanged, which requires $\Delta S = 0$ or $\pm 2$. However, the probability of approaching $\Delta S = \pm 2$ while maintaining the AFM spin ordering would be low, due to the frustration condition imposed by the triangular lattice. As a result, $\Delta S = 0$ is the most probable transition allowed within the AFM ordering as well as Raman selection rule. It would be possible to satisfy both conditions if all the three spins in one trimer rotate simultaneously by the same angle in the same direction. The symmetry of the triangular lattice permits only certain angles of rotation of the spins, namely, 60, 120, and 180 degrees. These three rotations are illustrated in Fig. 2b-d. Details of argument on the angular quantization of the spin-rotational excitation are given in the Supplementary Information. Number 1, 2, 3 spins are rotated counterclockwise from the spin structure of $\Gamma_4$ symmetry shown in Fig. 2a, by 60 degrees (Fig. 2b, green arrows), by 120 degrees (Fig. 2c, orange arrows), and by 180 degrees (Fig. 2d, red arrows), after excitation. Especially if all the three spins in one trimer are rotated 180 degrees (Fig. 2d), which is identical with the flipping of all three spins, the symmetry of the spin ordering would change from $\Gamma_4$ to $\Gamma_1$ representation locally. In analogy, these may be regarded as the $\Gamma_1$ bubbles in the $\Gamma_4$ sea.

A simple Hamiltonian is suggested below to address the spin excitations involving rotation of the spins in hexagonal LuMnO3 system. The Hamiltonian includes two terms for the spin interactions; first term is the spin-spin interaction between the Mn-ions within a trimer with spin exchange integral $J_1$ (intra-triangular interaction), and second is that between the Mn-ions in neighboring triangles with $J_2$ (inter-triangular interaction). Figure 3 shows a concept of the model (Fig. 3a) in the ground state ($\Gamma_4$) and that (Fig. 3b) in one of the excited states ($\Gamma_1$). There are six other trimers ($\Gamma_4$) around a norm trimer. Below $T_N$, Mn trimerization contracting in the Mn-ion *xy* plane is found in LuMnO3, which makes it possible to distinguish $J_1$ and $J_2$ [15, 19-21]. In Fig. 3a, number 1 spin has two nearest neighbors (number 2 and 3) connected with dotted line (intra-triangular interaction $J_1$) and four next nearest neighbors (number 2' and number 3') connected with dashed line (inter-triangular interaction $J_2$). Likewise, each of number 2 spin and number 3 spin has also two $J_1$ and four $J_2$ interactions. Considering double counting, the Hamiltonian should be as follows;

$$H = J_1 \left( \sum_{\substack{i,j=1 \\ (i \neq j)}}^{3} \vec{S}_i \cdot \vec{S}_j \right) + 2J_2 \left( \sum_{\substack{i,j=1 \\ (i \neq j)}}^{3} \vec{S}_i \cdot \vec{S}_{j\prime} \right) \qquad \text{Equation (1)}$$

where $\vec{S}_i$ is the $Mn^{3+}$-ion spin in one trimer, and $\vec{S}_{j\prime}$ is that in six neighboring trimers.

The largest energy difference from ground state is derived from the Hamiltonian when the number 1, 2, 3 spins are rotated 180 degrees, namely, three-spin flipping. Additionally, other notable energy values are corresponding to rotation of the spins by 60, and 120 degrees, and they are listed in table 1. First, it suggests energy differences, ΔE, calculated by the Equation (1). When the three spins in a trimer are rotated by the same angle simultaneously, there is no cost in the energy related to the stronger spin-spin interaction $J_1$. Thus ΔE is determined by the weaker interaction $J_2$ only. Calculation based on the model is performed taking total S = 2 of a $Mn^{3+}$-ion and assuming the excitation energy of the three-spin flipping is corresponding to the broad peak at ~ 805 $cm^{-1}$ in Fig. 1 [28]. From our model, we could obtain the value of $J_2 = 2.08$ meV, which is in reasonable agreement with various researches dealing with hexagonal $LuMnO_3$ [20, 33]. It is notable that the excitation energies due to three-spin rotation by 60 and 120 degrees precisely describe the broad peaks at 197 $cm^{-1}$ and 580 $cm^{-1}$ in Fig. 1, respectively.

Supplementary Figs. 2a, c, e, and g clearly show that spin rotations of a Mn-ion trimer by 0, 60, 120, and 180 degrees preserve the triangular symmetry by sustaining 30 or 60-degree angles between the spin directions. On the other hand, spin rotations by other angles, for example, 30, 90, and 150 degrees do not keep the triangular symmetry (Supplementary Figs. 2b, d, f). Consequently, only 60, 120, and 180-degree rotations of a trimer are allowed in the hexagonal crystal symmetry, and thus the spin excitation energies are quantized by the inherent triangular symmetry of the hexagonal $LuMnO_3$[36].

The spin excitation by 60, 120, and 180-degree rotation is a local excitation in one isolated trimer, not in the entire plane. These spin-rotational excitation of one Mn-ion trimer is like isolated flat bubble in the ground state of the $\Gamma_4$ sea. The spin symmetry of the bubble is locally different from the symmetry of the background.

A three-dimensional cartoon of the collection spin-rotational excitation bubbles is presented in Fig. 4 to help understand the nature of the isolated spin excitations. The spin ground states are abundant enough to form the $\Gamma_4$ sea below $T_N$. When a resonant light generating the Mn *d-d* transition is applied to a part of the $\Gamma_4$ sea, in-plane spin-rotational excitations would emerge with rotation angles of 60, 120, and 180 degrees. These spin excitations are isolated from each other, and each isolated excitation could be considered as an isolated flat bubble in the $\Gamma_4$ sea as pictured in Fig. 4. 60-degree rotations are depicted as green bubbles, 120-degree and 180-degree rotations, orange and red bubbles, respectively. Angularly-quantized spin-rotational excitation constitutes an example of energy quantization by the symmetry allowance.

**DISCUSSION**

We suggest a model based on a spin-spin interaction Hamiltonian to explain the spin excitation peaks observed in the Raman spectra of hexagonal LuMnO$_3$ in the cross configuration below T$_N$. Broad Raman peaks of hexagonal LuMnO$_3$ below T$_N$ are excited through the resonance with the Mn *d-d* transition by the incident red laser (~1.85 eV). Our model for the spin excitation suggests simultaneous rotation of the spins of the in-plane Mn$^{3+}$-ions to account for the energies of the Raman peaks. The model should meet several conditions: the Raman selection rule, preservation of the spin symmetry associated with the triangular lattice while maintaining the AFM spin ordering. A simple calculation is carried out to compare the model with our experimental Raman data. We could get a microscopic value for next nearest neighbor, $J_2 = 2.08$ meV, which is consistent with the results from the neutron scattering ($J_2 = 1.54$ meV [33]) and theoretical calculations ($J_2 = 2.37$ meV [20], 3 meV [37]). Based on the $J_2$ values obtained, $\Delta E$ values are calculated by the Hamiltonian equation (1), therefore, we could assign the broad peaks as the isolated spin excitations associated with the spin rotation by 60, 120, and 180 degrees. The ground spin state and the three-spin-flipping state represent $\Gamma_4$ and $\Gamma_1$ configuration, respectively, with the energy difference of ~ 0.1 eV (corresponding to ~805 cm$^{-1}$).

In this study, the spin excitation peaks of LuMnO$_3$ observed in the Raman scattering are due to the excitations solely in the Mn spins through the resonance with the Mn $d$-$d$ transition. Neutron scattering and magnetization measurements of $R$MnO$_3$ ($R$ = rare earths) on the other hand, are affected by the strong paramagnetic moment of the rare-earth ions, and the magnetic excitations by the Mn-ions are hard to differentiate from those related with the rare earths. That is, resonance Raman has a potential to differentiate the magnetic phase transition due to the Mn ions especially in hexagonal $R$MnO$_3$ system with strong paramagnetic moment of rare earth $R^{3+}$ ions other than Lu$^{3+}$ ions. Raman spectroscopy resonant with Mn $d$-$d$ transition suggests a good approach to study the spin ordering of the Mn ions in other hexagonal $R$MnO$_3$.

All the spin excitation peaks observed in LuMnO$_3$ below the Néel temperature by an inelastic light scattering are explained in terms of the Heisenberg spin-spin interaction Hamiltonian. We claim that the peaks at 197, 580, and 805 cm$^{-1}$ are due to excitations by Mn-ion spins rotated by 60, 120, and 180 degrees, respectively. The rotation angles are quantized by 60, 120, and 180 degrees, which is a consequence of the symmetry of the triangular lattice. The spin excitations are isolated in each triangular lattice. The isolated spin excitations may lead to optical control of the spin degree of freedom in future.

**MATERIALS AND METHODS**

Hexagonal LuMnO$_3$ single crystal was grown using the traveling floating zone method and characterized by magnetization, resistivity, and x-ray powder diffraction [38]. Platelet sample was cleaved perpendicular to the c axis. The sample area was 2.0 $mm \times 2.0\ mm$ with 0.2 $mm$ thickness. Helium-closed-cycle cryostat was used to control the temperature of the sample from 15 to 120 K in vacuum chamber. Raman scattering spectra were obtained by Horiba LabRam spectrometer coupled with a liquid-nitrogen-cooled CCD under $z(yx)\bar{z}$ cross configuration. Excitation light source was visible red laser which has continuous 671 nm (~1.85 eV) wavelength, with the power of 40 mW on the chamber window. Laser spot radius was about 0.8 $mm$ when using x40 objective lens. The background is subtracted from the raw Raman spectra and Adjacent-Average smoothing is performed by window size 7, threshold 0.05 after the subtraction. Whole data are normalized by the $A_1$ phonon (~680 cm$^{-1}$) intensity and we considered temperature dependence of the $A_1$ phonon [12] for each temperature when normalizing the spectra.

## ACKNOWLEDGEMENT

The measurements in this work were supported by Korea Basic Science Institute (National Research Facilities and Equipment Center) grant funded by the Ministry of Education (2020R 1A 6C 101B194). Part of this research was supported by Basic Science Research Program through the National Research Foundation of Korea (NRF) funded by the Ministry of Education (2021R1A6A1A10039823) I. S. Y. acknowledges the financial support by the National Research Foundation of Korea (NRF) grant funded by the Korean government (Grant No. 2017R1A2B2009309). The work at Rutgers University was supported by the DOE under Grant No. DOE: DE-FG02-07ER46382.

## AUTHOR CONTRIBUTIONS

S. Kim performed the experiments, carried out the analyses of data, and designed a model to explain the origin of the spin-excitation. S. Kim and J. Nam contributed to the experimental set-up and data processing. X. Xu and S.-W. Cheong synthesized and characterized the crystals. I.-S. Yang conceived the project and supervised the research. S. Kim and I.-S. Yang wrote the draft and revised the manuscript. All authors have discussed the results and reviewed the manuscript.

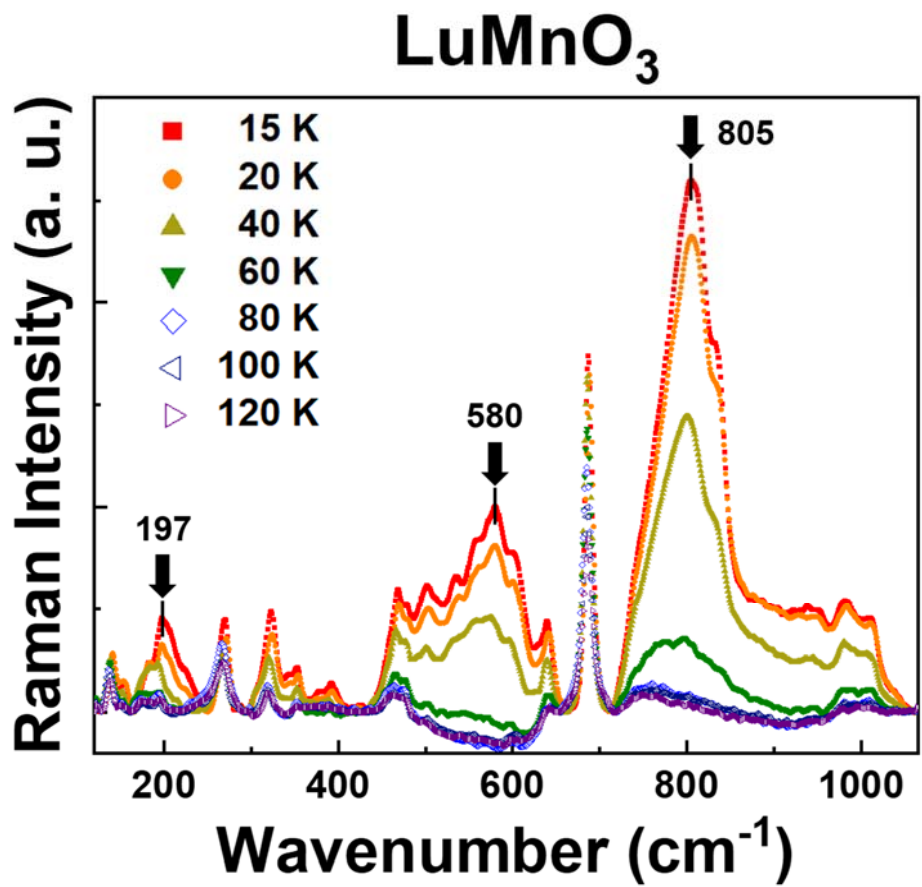

Figure 1. Temperature dependence of Raman spectra of hexagonal LuMnO$_3$ single crystal in the cross scattering configuration. Several broad peaks (~197, 580, 805 cm$^{-1}$) are prominent at low temperatures, and disappear above 80 K (~T$_N$).

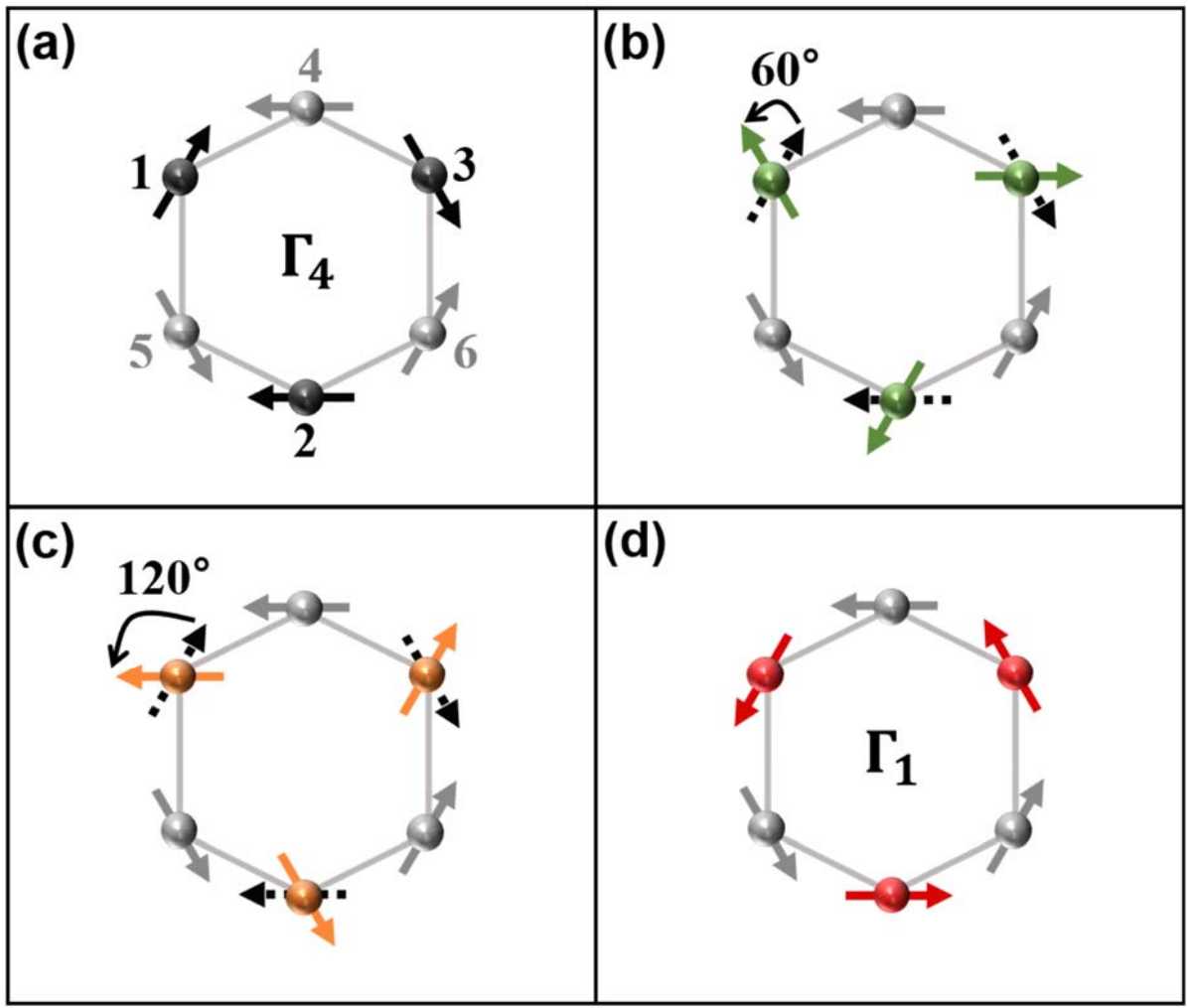

Figure 2. Several AFM spin configurations of hexagonal LuMnO$_3$ single crystal. Mn-ions are indicated by solid spheres: gray spheres (number 4, 5, 6) are located in the $z = \frac{c}{2}$ plane and black or colored ones (number 1, 2, 3) are in the $z = 0$ plane. Each arrow overlapped with the sphere expresses the spin direction of a Mn$^{3+}$-ion (S = 2). We assume that colored spheres only are excited by the incident light. (a) The $\Gamma_4$ spin structure in the ground state. (b) Three spins (1, 2, 3) rotated counterclockwise by 60 degrees (green arrows), or (c) by 120 degrees (orange arrows). (d) Three spins (1, 2, 3) are flipped from the $\Gamma_4$ structure, resulting in the $\Gamma_1$ spin structure locally (red arrows).

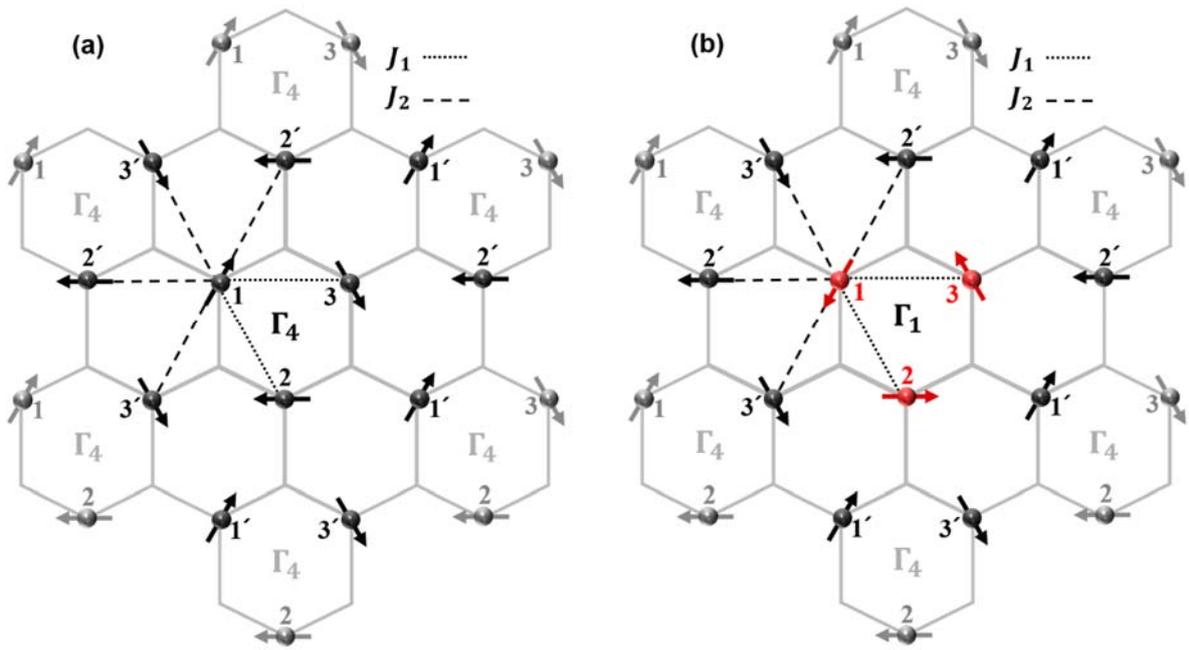

Figure 3. A trimer composed of three Mn-ions (center), and its six neighbors in the $z = 0$ plane in hexagonal LuMnO$_3$ single crystal. Below T$_N$, each trimer has two kinds of spin exchange integral: Dotted lines (nearest neighbor) indicate spin exchange integral $J_1$, and dashed lines (next nearest neighbor) indicate spin exchange integral $J_2$. (a) Every trimer has $\Gamma_4$ spin structure in the ground state. (b) Norm trimer has the $\Gamma_1$ spin structure locally after the spin-flip excitation, while its six neighbors remain in the $\Gamma_4$ spin structure.

| Rotation angle | ΔE | Model Calculation | | Experimental Data | |
|---|---|---|---|---|---|
| | | ( meV ) | ( cm$^{-1}$ ) | ( meV ) | ( cm$^{-1}$ ) |
| 60° | $3S^2 J_2$ | 24.95 | 201.3 | 24.43 | 197 |
| 120° | $9S^2 J_2$ | 74.85 | 603.7 | 71.91 | 580 |
| 180° | $12S^2 J_2$ | 99.80 | 804.9 | 99.81 | 805 |

Table 1. Excitation energy values calculated from the model Hamiltonian in the text corresponding to the rotation angles of the three Mn-spins in a trimer in LuMnO$_3$ single crystal. In the model, $S = 2$ and $J_2 = 2.08$ meV (Park, J. *et al*, Oh, J. *et al* ) are assumed.

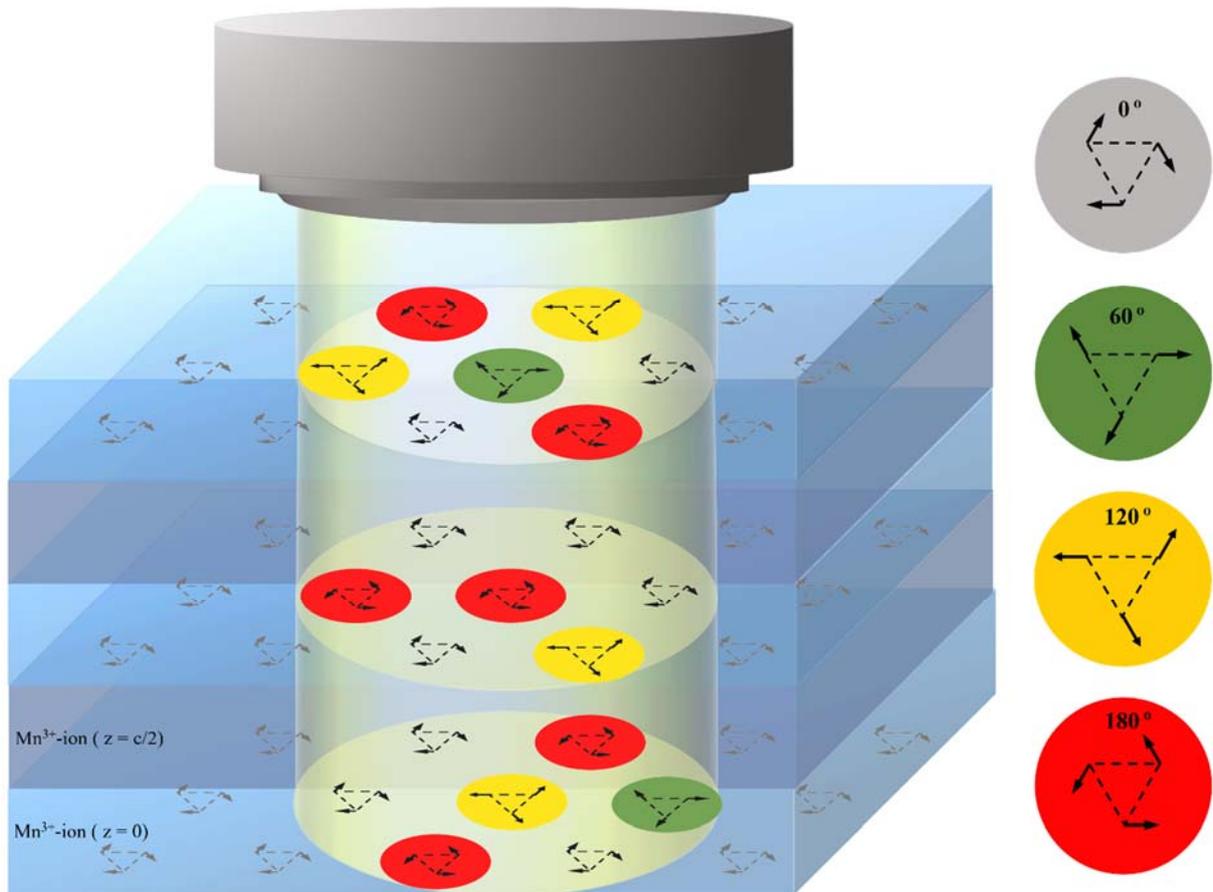

Figure 4. A visualization of the isolated spin-rotational excitations. Below $T_N$, the $Mn^{3+}$-ions are lying on each *xy* plane maintaining the $\Gamma_4$ spin structure. The spin states are excited by simultaneous rotation of the three spins in a trimer by 60 degrees (green bubbles), 120 degrees (orange bubbles), and by 180 degrees (red bubbles) under the resonant light. The symmetry of the triangular lattice allows only these three rotations in hexagonal $LuMnO_3$.